\documentclass[12pt]{article}

\usepackage{graphics}
\usepackage[dvips]{color}
\usepackage{multicol}

\usepackage{amsmath}
\definecolor{white}{rgb}{1,1,1}
\definecolor{yellow}{rgb}{0.95,0.75,0.1}
\definecolor{red}{rgb}{0.5,0,0}
\definecolor{green}{rgb}{0,1,0}
\definecolor{blue}{rgb}{0,0.5,1}
\definecolor{bgcolor}{rgb}{0.94,0.91,0.78}

\definecolor{lblue}{rgb}{0,0.8,1}
\definecolor{dblue}{rgb}{0,0,.6}
\definecolor{dgreen}{rgb}{0,0.3,0}
\definecolor{lila}{rgb}{0.8,0,0.8}
\definecolor{violet}{rgb}{1,0,1}
\definecolor{grey}{rgb}{0.3,0.3,0.3}
\definecolor{turquoise}{rgb}{0,.9608,1}

\def\lsim{\raise0.3ex\hbox{$\;<$\kern-0.75em\raise-1.1ex\hbox{$\sim\;$}}}
\def\gsim{\raise0.3ex\hbox{$\;>$\kern-0.75em\raise-1.1ex\hbox{$\sim\;$}}}

\newcommand{\nee}{\nonumber \end{eqnarray}} 
 
\usepackage[dvips]{color}
\usepackage{multicol}
\usepackage{amsmath,color,graphics,graphpap}
\def\lsim{\raise0.3ex\hbox{$\;<$\kern-0.75em\raise-1.1ex\hbox{$\sim\;$}}}
\def\gsim{\raise0.3ex\hbox{$\;>$\kern-0.75em\raise-1.1ex\hbox{$\sim\;$}}}

\usepackage{amsmath,color,graphics,graphpap}
\newcommand{\be}{\begin{eqnarray}}
\newcommand{\ben}{\begin{eqnarray}\nonumber}
\newcommand{\ee}{\end{eqnarray}}

\renewcommand{\theequation}{\thesection.\arabic{equation}}
\renewcommand{\thesection}{\arabic{section}}
\definecolor{lred}{rgb}{1,0.3,0.3}
\definecolor{red}{rgb}{0.5,0,0}
\definecolor{dblue}{rgb}{0,0,.6}

\def\ApJ{{ Astrophys. J.} }

\def\JCAP{{\it Journ. of Cosmol. \& Astropart. Phys.}Ê}

\def\MNRAS{{ Month. Not. Roy. Astr. Soc.} }

\def\PRD{{ Phys. Rev.} {\bf D} }

\usepackage{amsmath}
\usepackage{amsmath,color,fancybox,graphics,graphpap,rotating}
\begin{document}

\title {Extending the susy model to core-collapse supernovae}
\author{ L.
Clavelli\footnote{Louis.Clavelli@Tufts.edu,\,lclavell@ua.edu}\\ 
Dept. of Physics and Astronomy, Tufts University, Medford MA 02155\\ 
Dept. of Physics and Astronomy, Univ. of Alabama, Tuscaloosa AL 35487}
\date{Aug 25, 2019}
\maketitle

\begin{abstract}
A noteable feature of the two standard models for thermonuclear and
core-collapse supernovae is that, although these two models are fundamentally
different, the respective supernova types have quite similar rates and
appearances.  For instance, both types occur one to several times per century
per typical galaxy and both types seed the universe with the heavy elements
essential to life.  In spite of this,
neither standard model provides a reasonably problem-free description of its target 
phenomenon. A major obstacle
to providing a unified picture of supernovae would seem to be the fact
that type Ia supernovae occur typically with gigayear delay times after the
cessation of carbon fusion while the core-collapse explosions occur
only days after such fusion cessation. 
In this article we study the possibility of extending the successful
supersymmetric model for type Ia supernovae to core-collapse events. 
The question is whether and under what assumptions a phase transition to
an exact supersymmetric background can efficiently explain both type Ia 
and core collapse supernovae.
\end{abstract}   

\section{Introduction}

The standard thinking about Type Ia supernovae, originating more than forty
years ago, is that mass transfer from a binary partner to a white dwarf star 
induces fusion at or near
the Chandrasekhar mass limit. For a review see \cite{Maoz-Mannucci-Nelemans}. 
An alternative 
model \cite{Biermann-Clavelli},\cite{SNIa} for  
Type Ia supernovae 
is based on the somewhat radical hypothesis that matter at sufficiently high 
density undergoes a phase transition to a background of exact supersymmetry (susy).  
In such
a background, generically available mechanisms exist to convert pairs of fermions 
into pairs of degenerate bosons which, due to their Pauli Principle immunity,
can deposit large amounts of degeneracy energy into the surrounding matter. 
Although susy particles have not as yet been discovered in the broken susy phase, 
It has been shown that this model avoids \cite{SixIndications} six major 
puzzles of the 
standard binary models for Type Ia events and easily fits the delay time 
distribution and other observations.  
The model also provides some understanding of the Phillips Relation which is 
key to the cosmological usefulness of Type Ia supernovae and explains
most of the host-galaxy effect \cite{galfx} which affects the supernova 
measurement of the Hubble Constant.

The standard model for Type II and other core-collapse supernovae on the other hand  
postulates that the outer shell of a dense star is propelled to large
distances by bouncing off a dense inner core and/or receiving an impulse from
a neutrino cloud \cite{Bethe}.  However, in a large number of detailed monte carlos the 
explosion stalls out and the neutrino interactions are too weak to re-ignite
the supernova \cite{Thielemann}.  This situation has lead to a continuing search for initial 
conditions that will necessarily lead to a core collapse explosion. (An August 2019 
search for titles including "progenitor" yielded 1008 papers.)
Obviously, a very particular initial condition requirement will tend to 
reduce the predicted supernova rate.  
A correlated problem is that the standard model simulations 
fail to produce the heaviest element abundances observed in the universe.   
The observation of gravitational waves from binary neutron stars led to some
hope that such mergers would explain the existence of heavy elements but
further study has made it clear that
these are also not able to fully account for the
abundances \cite{Frebel},\cite{Haynes},\cite{Siegel}.  Attention has therefore 
been returned to core collapse
supernovae as a heavy nuclei source.
It is generally admitted that some additional energy release from beyond
the standard model would be helpful.  In the susy model the extra energy 
released per unit mass is defined by the degeneracy energy in the progenitor
atoms.

  In this situation it is natural to ask whether the susy model for type Ia
supernovae can be extended to type II and other core-collapse events.
The main obstacle to such a unification would seem to be the great difference
in time scales as mentioned in the abstract above.
   
In Section\,\ref{sec-delay}, we discuss the time delays for both kinds
of supernovae from the point of view of the susy model.
This is followed by further sections on bubble nucleation, on
fusion stages in massive stars, on heavy nuclei production, on neutron star 
remnants, and on the black hole gap.
Some summary is presented in the final section.

\section{supernova delay time}
\label{sec-delay}

Assuming the ground state of the universe is supersymmetric, the transition
probability per unit time to exact susy based on the bubble nucleation formalism
\cite{Coleman} enhanced in dense matter as laid out in 
\cite{Biermann-Clavelli},\,\cite{SNIa},\, \cite{SixIndications},\,\cite{galfx} is 
 
\be \frac{dP}{dt} = A\, \int d^3r\,e^{-B(r)} 
\label{TransRate}
\ee

where, in dense matter, the action, $B(r)$, is inversely proportional to the cube of the 
degeneracy energy which is, itself, proportional to the matter density.
\be
     B(r) = (\frac{\rho_c}{Z\,\rho(r)})^3\quad .
\label{action}
\ee

$A$ and $\rho_c$ are at present free parameters.  In the next section, it is 
argued that the transition probability per unit space-time volume 
should continue to increase at high density suggesting that a third
parameter must be added in the extension of the model to core collapse supernova. 

In the absence of such an extension, the fit to the SN Ia delay time 
distribution \cite{galfx} gives
\be\nonumber
   A &=& 30.0\, R_E^{-3} \mathrm{Gyr}^{-1}\\
   \rho_c &=& 58.8 M_\odot /R_E^3  \,\quad .
\label{params}
\ee
The density conversion from solar system units (solar mass, $M_\odot$, and 
earth radius, $R_E$,) to cgs units is given by
\be
    M_\odot \, R_E^{-3} = 7.661\,10^6 \,\mathrm{g/cm^3} \quad .
\ee
Since protons and neutrons have separate Pauli towers, the degeneracy energy is
proportional to the average 
of the proton and neutron number denoted here by $Z$.  
In low lying dominant nuclei (up to Silicon) this is equal to the 
atomic number.

As can be seen from \ref{TransRate} the system has an inverse lifetime in the quantum
mechanical sense given by
\be
     \tau^{-1} = A\,  V_\mathrm{eff}
\label{tau}
\ee
where the effective volume is 
\be  V_\mathrm{eff}=\int{d^3r}\,e^{-B(r)} 
\quad .
\label{EffVol}
\ee 

\section{Beyond the Coleman-DeLuccia approximation}
\label{BeyondCdL}

    The transition rate to an exact susy background, as given by 
eq.\,\ref{TransRate} and eq.\,\ref{action} is based on the 
Coleman formula \cite{Coleman} which in turn is based on
the assumption that, in the thin wall approximation, the path integral is dominated by the
path of minimum action.  The assumption is that this one path dominates over the sum of a possibly
large number of other nearby paths.  Corrections to the resulting
expressions must be expected especially in the case when the action is not sufficiently large.  In \cite{SNIa}, we considered relatively positive corrections to eq.\ref{action} 
in a minimum $\chi^2$ fit to the delay time distribution (DTD) of type Ia supernova events.
On the other hand, one could argue that the first corrections should be relatively negative.
This could follow from the observation that the action in eq.\ref{action} saturates at
large density whereas one might expect that the transition rate continues to increase 
rapidly at increasing density.  We write, therefore, as in \cite{SNIa}
\be
      B(r) = (\frac{\rho_c}{Z\,\rho(r)})^3 - b_0\, Z\,\rho(r)/\rho_c \quad 
\label{actionMod}
\ee
but here we seek fits with a relatively negative correction to the Coleman
formula, (i.e. positive $b_0$ in eq.\,\ref{actionMod}).
We ask whether, with free parameters A and $\rho_c$ of eqs.\ref{TransRate} and \ref{action},
and the $b_0$ of eq.\,\ref{actionMod}, we can:
{\flushleft 1) find a good fit to the delay time distribution of type Ia supernovae\,
and simultaneously}
{\flushleft 2) explain the time delay, rate, energy release, and heavy nuclei production in core-collapse events.}

If $b_0$ is too small, the lifetime of massive stars after fusion cessation will, like white dwarfs, be at or near Gigayr time scales contrary to observation.
If $b_0$ is too large the previous successful fits to the SN Ia delay time distribution will be unacceptably disturbed.  

Observations of neutron stars also have the potential to prevent
the extension of the susy model to core-collapse events.  Unless $b_0$ is microscopic
or negative the neutron star lifetime prediction could be unacceptably short.  On the other hand, the susy model with positive $b_0$ could positively 
impact the theory of what we call neutron stars the current theory of 
which is not without puzzles.  This is discussed below in section\,\ref{sec-remnant}.    

We begin, therefore, by finding values of 
$b_0$ which yield acceptable $\chi^2$ fits to the SN Ia delay time distribution.  The phase transition model depends on the reasonably well known single white dwarf production rate as a function of birth mass distribution. 
\be
    F(M) = a_0 \,((M/M_\odot-0.478)/0.09)^{-2.35} \quad ,
\label{imf}
\ee
where 
\be
     a_0 = 5.48\,\mathrm{yr^{-1} gal^{-1}}
\ee
and the typical galaxy, gal, is defined to be a set of
$10^{10}$ naturally produced solar mass stars.

The fit to the mass distribution of a sample of hot white dwarfs is shown 
in ref.\,\cite{SNIa}.  This sample is from relatively nearby white dwarfs from 
correspondingly moderate metellicity regions while
the observed supernovae sample a wider range of metallicities as discussed in
ref.\,\cite{galfx}. 
The SN Ia delay time distribution is then
\be
    \frac{dN}{dt} = \int dM\,F(M)\, e^{-t/\tau(M)}/\tau(M) \quad .
    \label{DTD} 
\ee   
For small values of $b_0$, and values of the other two parameters near eq\,\ref{params}, we calculate $\chi^2$ for the three most accurate DTD values
of ref.\,\cite{MaozMannucciBrandt}:
\be\nonumber
     t_1 &=& 0.039 Gyr  \quad dN/dt = (0.0138 ^{+0.0032} _{-0.0029}) Gyr^{-1}\\\nonumber
     t_2 &=& 0.416 Gyr  \quad dN/dt = (0.00256 ^{+0.00058}_{-0.00066}) Gyr^{-1}\\
     t_3 &=& 2.376 Gyr  \quad dN/dt = (0.000183 ^{+0.0000359}_{-0.000041}) Gyr{-1}\nonumber  
\ee

\begin{table}[ht]
\begin{center}
\begin{tabular}{|c|c||c|c|c|}\hline
       $b_0$   &\quad\, $\chi^2$ \, & \quad $\rho_c$\, ($M_\odot/R_E^3$)\, & \quad $\tau_0 \,(\mathrm{Gyr})$\,&A\, ($\mathrm{Gyr}^{-1}{R_E}^{-3}$) \cr\hline
       0.00  &\,  0.002 \,  &\,  58.8 \, & \, 0.418 &\,30.0 \cr\hline
       0.01  &\,  0.006	\,  &\,  58.8 \, & \, 0.420 &\,28.9 \cr\hline
       0.02  &\,  0.016 \,  &\,  58.8 \, & \, 0.422 &\,27.8 \cr\hline
       0.03  &\,  7.159	\,  &\,  59.8 \, & \, 0.414 &\,12.7 \cr\hline
\end{tabular}      
\caption{Minimum $\chi^2$ fits to the Type Ia supernova delay time distribution for various small values of $b_0$ with $\rho_c$ and $\tau_0$ values near the zero $b_0$ fit. 
Values of $b_0$ less than or equal to $0.02$ give excellent fits to the data.  }
\label{table0}
\end{center}
\end{table}

\begin{figure}[ht]
\centering
\includegraphics[scale=0.65]{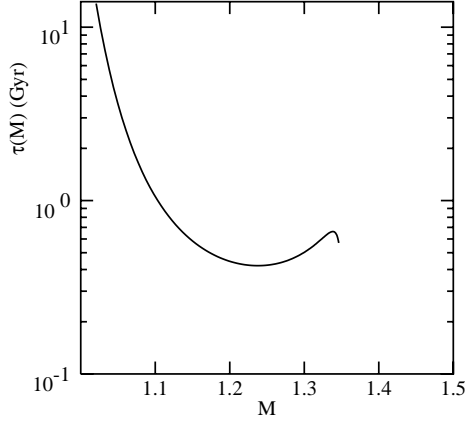}
\caption{
The lifetime, $\tau(M)$, vs white dwarf mass, $M$, with $b_0=0.02$.
Unlike the $b_0=0$ plot shown in \cite{SNIa}, with the modified action the
lifetime turns down slightly at large $M$. Of course at $M>1.38$ the lifetime
would essentially go to zero due to normal fusion.
}
\label{tauplot}
\end{figure}

\begin{figure}[ht]
\centering
\includegraphics[scale=0.65]{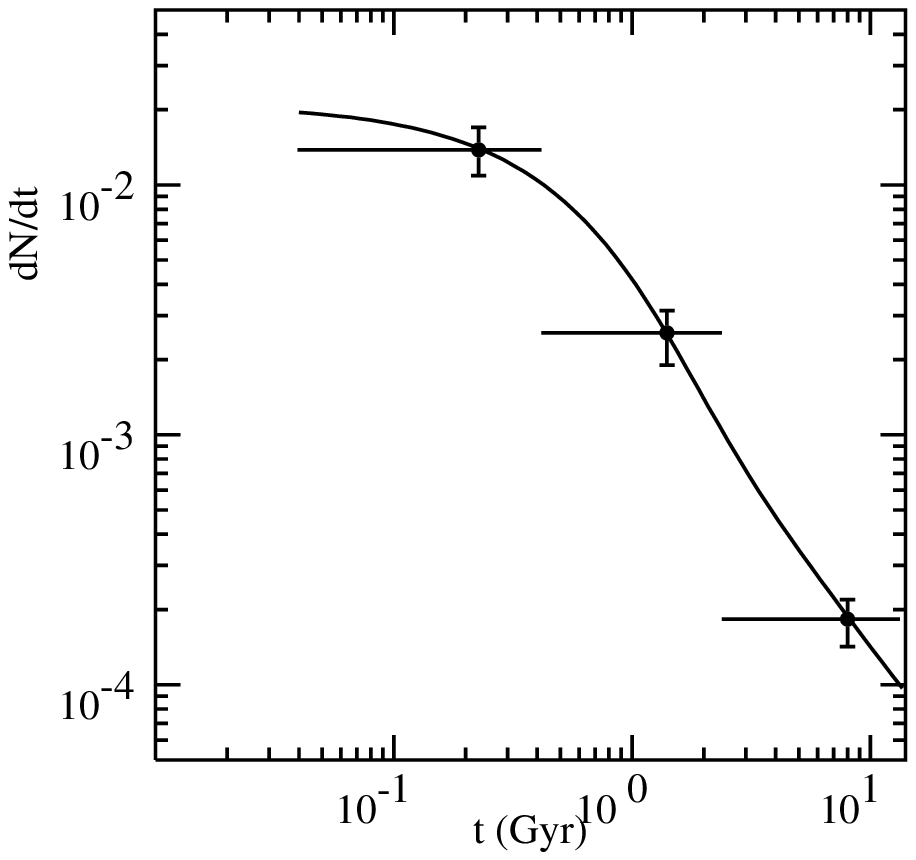}
\caption{The best fit to the delay time distribution with fixed $b_0=0.02$.  
The fit is not visibly worse than the unmodified ($b_0=0$) Coleman fit
shown in ref.\,\cite{SNIa}.}
\label{DTDplot}
\end{figure}

It cannot be ruled out that other, more distant values of the parameters, also give good fits but, certainly, they cannot be statistically much preferred over the excellent $b_0=0.02$ fit.  In this work we fix $b_0=0.02$ and study the effect on neutron star theory and core-collapse supernovae.  Technically, the $b_0 =0$ fit has a lower $\chi^2$ but its improvement over the $b_0=0.02$ fit is statistically insignificant. 
The plot of $\tau$ vs white dwarf mass, $M$, with $b_0=0.02$ is shown in 
fig.\ref{tauplot} and the delay time distribution for SN Ia events with the chosen $b_0$ value is shown in fig.\,\ref{DTDplot}.

\section{Likelihood of susy transition during stages of fusion and gravitational collapse}
\label{stages}

Unless interrupted by a phase transition, a massive star will undergo successive
stages of fusion of heavier elements with decreasing energy output.  In 
table \ref{table2} we consider 
fusion in a thirty solar mass star with a three solar mass core following \cite{Heger}.  Because
the density during fusion is relatively low, the probability of a phase transition
following eq.\,\ref{TransRate} is negligible.  In the silicon phase we use $Z=28$ in
eq.\,\ref{actionMod} since the transition in the core is enhanced by the significantly higher degeneracy energy of the accumulating Ni$^{56}$.

For more massive stars the 
fusion duration is much shorter making the likelihood of 
the phase transition even more negligible.  As can be seen from table\,\ref{table2}
during the final fusion stage the lifetime against the susy phase transition is
considerably shortened from the Gyr scale but is still much too long to 
make the transition likely.  
It remains to consider the 
likelihood of the susy transition during the collapse phase after cessation of fusion. 

\begin{table}[ht]
\begin{center}
\begin{tabular}{|c|c|c|c|}\hline
   stage    & $\quad \rho$\, (g/cm$^3$)\, & duration\, (yr) & \quad $\tau$ \,(yr)\, \cr\hline
   neon burning to  O, Mg &\,  $10^9$ \, & 0.6 & \, $1.9\cdot 10^{84}$ \, \cr\hline
   oxygen burning to  Si, S  &\, $10^9$ \, & 2.0 & \, $2.5\cdot 10^{21}$ \, \cr\hline
   silicon burning to  Ni$^{56}$ &\, $10^{10}$ \,& 0.031  & \, $3.8\cdot 10^4$ \, \cr\hline
\end{tabular}      
\caption{ density, and duration of successive states of fusion 
for a 30 solar mass star \cite{Heger} compared
with calculated lifetime, $\tau$, before susy phase transition.  Since, in all stages, the 
lifetime before phase transition is much greater than the duration of fusion 
the likelihood of phase transition during burning is negligible.}
\label{table2}
\end{center}
\end{table}
%
%

After the energy release from fusion becomes too weak to support the star
it undergoes a rapid collapse.  The transition rate to the exact susy ground
state during this phase is as given in eq.\,
\ref{TransRate} but the density is now a function of both space and time.
\be
     \frac{dP}{dt} = A \int d^3\,r \,e^{-B(\rho(r,t))}\quad .
\ee
The probability of a phase transition between times $t_1$ and $t_2$ is
\be
      P = A \int_{t_1}^{t_2}\,dt\, \int d^3\,r \, e^{-B(\rho(r,t))}\quad .
\ee
If $t_1$ and $t_2$ are separated by a small $\Delta t$, this probability,
using eq.\,\ref{tau}, is
\be
      \Delta P = A\,\Delta t\,\tau^{-1}(t) \quad .
\ee
If $\tau(t)$ is significantly less than $\Delta t$, the susy phase transition
is preferred.  The enormous degeneracy energy that is then released is much 
greater than the energy released in silicon fusion and efficiently
explodes the star creating copious heavy elements in the ejecta. 

We approximate $\tau^{-1}(t)$ by 
\be
      \tau^{-1} = \frac{A\,M_{\mathrm{core}}}{\rho_{\mathrm{peak}}}\, e^{-B(\rho_{\mathrm{peak}}(t))}
\label{TauEff}
\ee
where $\rho_\mathrm{peak}$ is
the average density of a core of mass $M_\mathrm{core}$ at time $t$.  
The behavior of the lifetime
is dominated by the exponential behavior and is less sensitive to the 
pre-factor.

During the final silicon fusion stage, the star expands and comes 
momentarily to rest at a low density 
after which gravitational collapse begins. We take this
initial density, somewhat arbitrarily, to be one tenth of the nominal 
white dwarf density 
\be
    \rho_\mathrm{ini} = \frac{3 M_\odot}{40\, \pi {R_E}^3} = 1.83\cdot 10^5\,\mathrm{g/cm^3} \quad .
\ee
At this time, $t=0$, the star, especially in the core, is largely iron
so any further exothermic fusion is negligible.
We consider five spherically symmetric matter distributions (stars)
with masses
\be
    M(n) = M_\odot \cdot 10^n \qquad n=1,...,5 \quad .
\ee
During the collapse we define $100$ milestone mean densities, the last
being the mean density of a black hole of mass $M(n)$.
\be
     \rho(i) = \rho_\mathrm{ini} \cdot e^{\frac{i-1}{99} \ln (\rho_{BH}(n)/\rho_\mathrm{ini})} 
    \qquad i = 1,...,100 \quad . 
\ee
The mean density of a black hole of mass M is
\be
     \rho_{BH} = \frac{3 M(n)}{4\,\pi\,{R_S}^3} = \frac{3 c^6}{32\pi {G_N}^3 {M(n)}^2} \quad .
\ee
We consider a small differential of mass, $\delta m$ near the stellar edge, at radius $r$ from the center,
\be
   r = (\frac{3 M(n)}{4 \pi \rho(r)})^{1/3} \quad .
\label{radius}
\ee
Below the Chandrasekhar mass, $M_c \approx 1.4\,M_\odot$, there is a stable
(time independent) solution.  Above this mass, since further fusion energy release
becomes negligible, the radius r decreases with $M(n)$ fixed and the momentum of the
element of mass, $\delta m$, given relativistically by
\be
   p = \frac{\delta m\, v}{\sqrt{1-v^2/c^2}} \quad ,
\ee
increases inwardly at a rate given by the sum of the gravitational and
electron degeneracy forces \cite{Chandra}
\be
  F_G + F_D = c^2 \delta m \frac{d}{dr} (\frac{G_N M(n)}{r c^2} + a b^3 \sqrt {1+ b^2 \rho^{2/3}}) \quad ,
\ee
where, in terms of electron and nucleon masses,
\be 
     a &=& \frac{{m_e}^4}{3 \pi^2 {(\hbar c)}^3} \\
     b &=& \frac{\hbar c}{m_e}{(\frac{3 \pi^2}{2 M_N})}^{1/3} \quad .
\ee
Supplying the integration constant by requiring that $v=0$ at $\rho=0$ and
using eq.\,\ref{radius} leads to 
\be
     {(1-v^2/c^2)}^{-1/2} = Y \quad .
\ee
with
\be
       Y = 1 + \frac{G_N}{c^2} {M(n)}^{2/3} {(4 \pi \rho/3)}^{1/3}
        - a b^3 (\sqrt{1+b^2 \rho^{2/3}} -1) \quad .
\ee
Thus
\be
      v = \frac{dr}{dt} = c {\sqrt {Y^2 -1}}/Y
\ee
or, using eq.\,\ref{radius}, the time between the $i'th$ and $(i+1)'th$ stages is 
\be
   \Delta t(i) = (\frac{3 M(n)}{4 \pi})^{1/3} \frac{Y}{{\sqrt{Y^2 -1}}} \Delta (\rho^{-1/3}) \quad .
\ee

In the non-relativistic limit appropriate to lower mass progenitors, this is 
\be
     \Delta t(i) = \frac{\Delta (\rho^{-1/2})}{\sqrt{6 \pi G_N}} (1 - 
\frac{c^2\,a\,b^4}{G_N\,M(n)}(\frac{3 M(n)}{4 \pi})^{1/3})^{-1/2} \quad .
\ee

The phase transition interrupts the collapse if and when $\Delta t(i)$ is significantly 
greater 
than the lifetime, $\tau(i)$, from eq.\,\ref{tau} with the modified action from 
eq.\,\ref{actionMod}.  The density during collapse is strongly peaked at the origin.
Without committing to their suggestion of an energy release mechanism, one can
note the modeling of the density peaking  in ref.\,\cite{Feng}.

We tentatively assume from Chandrasekhar's treatment of gaseous stars \cite{Chandra}
that the peak density, which governs the bubble nucleation, is some 54 times greater
than the average density. 
As a rough approximation, we write, therefore,
\be
      \rho_{\mathrm {peak}} &=& 54\, \rho\\
      M_{\mathrm core} &=& M(n)/10 \quad .
\ee
Our results are not qualitatively different if we vary the numerical parameters here by factors of order unity.

In advance of a detailed calculation of the integral over the density profile, we
write at the $i'th$ stage in the collapse
\be
     \tau(i) = \frac{\rho_{\mathrm{peak}}}{A M_{\mathrm{core}}} e^{B(\rho_{\mathrm peak})}
\ee
with
\be
     B_{\mathrm{peak}} = \frac{\rho_c}{54\, Z\, \rho} - 0.02 \frac{54\, Z\, \rho}{\rho_c} \quad .
\ee
Here, we take $Z=28$ corresponding to one half the atomic weight of the iron core
since the degeneracy energy in iron comes from both protons and neutrons.
With the $0.02$ correction to the bare Coleman formula the lifetime goes
exponentially to zero at large density.  The results are given in table\,\ref{table3}.

\begin{table}[ht]
\begin{center}
\begin{tabular}{||c|c|c|c|c|c|c|c||}\hline
\, n &\, i \,&\,  $\rho\;$ (g/cm$^3$) \,&\,  t\;(s)\,&\, v/c \,&\, $\Delta t\;$ (s) \,&\,$\tau\;$  (s)\,&\, $\tau/{\Delta t}$ \, \cr\hline\hline
\, 1 \,&\, 40 \,&\,6.4$\cdot 10^{8}$\,&\,0.68\,&\,0.11 \,&\,1.2$\cdot 10^{-3}$\,&\,8.0$\cdot 10^{-1}$\,&\, 6.3$\cdot 10^{2}$ \cr \hline
\, 1 \,&\, 41 \,&\,7.9$\cdot 10^{8}$\,&\,0.68\,&\,0.11 \,&\,1.1$\cdot 10^{-3}$\,&\,4.1$\cdot 10^{-5}$\,&\, 3.6$\cdot 10^{-2}$ \cr \hline
\hline
\, 2 \,&\, 51 \,&\,6.3$\cdot 10^{8}$\,&\,0.66\,&\,0.25 \,&\,9.2$\cdot 10^{-4}$\,&\,2.0$\cdot 10^{-1}$\,&\, 2.2$\cdot 10^{2}$ \cr \hline
\, 2 \,&\, 52 \,&\,7.4$\cdot 10^{8}$\,&\,0.66\,&\,0.26 \,&\,8.5$\cdot 10^{-4}$\,&\,1.3$\cdot 10^{-4}$\,&\, 1.5$\cdot 10^{-1}$ \cr \hline
\hline
\, 3 \,&\, 71 \,&\,6.3$\cdot 10^{8}$\,&\,0.66\,&\,0.51 \,&\,7.1$\cdot 10^{-4}$\,&\,1.8$\cdot 10^{-2}$\,&\, 2.6$\cdot 10^{1}$ \cr \hline
\, 3 \,&\, 72 \,&\,7.0$\cdot 10^{8}$\,&\,0.66\,&\,0.52 \,&\,6.7$\cdot 10^{-4}$\,&\,1.1$\cdot 10^{-4}$\,&\, 1.6$\cdot 10^{-1}$ \cr \hline
\hline
\, 4 \,&\, 99 \,&\,1.7$\cdot 10^{8}$\,&\,0.69\,&\,0.74 \,&\,9.8$\cdot 10^{-4}$\,&\,1.3$\cdot 10^{10}$\,&\, 1.3$\cdot 10^{13}$ \cr \hline
\hline
\, 5 \,&\, 99 \,&\,1.8$\cdot 10^{6}$\,&\,0.55\,&\,0.74 \,&\,3.2$\cdot 10^{-3}$\,&\,1.2$\cdot 10^{12}$\,&\, 3.9$\cdot 10^{14}$ \cr \hline
\hline
\end{tabular}
\caption{various quantities are tabulated here during
the collapse including the mean density, $\rho$, the
elapsed time, t, since initial density, the collapse 
speed, $v/c$, the time interval between mileposts, the
lifetime against phase transition, and their ratio. 
For progenitor masses less than some $10^4\,M_\odot$
the lifetime against phase transition becomes suddenly
less than the time step when the mean density exceeds
about $7 \cdot 10^8$ \,g/cm$^3$.}
\label{table3}
\end{center}
\end{table}

  For stars up to mass $10^3 M_\odot$ table\,\ref{table3} shows
there is a last stage at which $\tau/\Delta t$ is greater than unity meaning
the phase transition is unlikely followed by a sharp transition to a stage
where it is negative implying a rapid phase transition. For masses
$10^4 M_\odot$ and $10^5 M_\odot$, the ratio remains greater than unity 
meaning that the star collapses to a black hole without becoming supernova.
Further discussion of the possibility that core collapse supernovae are
due to a transition to the exact susy phase are reserved for section\,\ref{sec-summary}.


\section{Heavy nuclei production}
\label{sec-heavy}

The production of very heavy elements is a long-standing astrophysical problem
since fusion alone cannot produce copious amounts of elements above iron.
The successive stages of nuclear fusion in a massive star as discussed in 
section \ref{stages} naturally lead to a star totally or partially devoid of
elements above iron which is then subject to gravitational collapse.  
At sufficiently high density the internal kinetic energy can dissociate 
iron into its constituents but an external energy source is still necessary
to accelerate them above escape velocity.  Heavy elements can then be produced
in the ejecta through the rapid neutron capture process on remaining iron
nuclei.  The required energy input is the same as can be calculated from
the direct endothermic fusion of iron into the heavier elements as discussed 
below.    

As a star contracts its potential energy becomes 
more negative and the kinetic energy of its constituents increases but its
binding energy also increases.  Thus, as was known to Newton, it becomes 
increasingly difficult to throw out any significant fraction of its mass.
Even with Einstein's extension it is known that gravity by 
itself cannot result in the ejection of large amounts of matter from an 
isolated star. Furthermore, all other standard model interactions on iron 
are endothermic and cannot trigger a supernova explosion.  
Turbulence \cite{Melson} and 
asymmetric fluctuations might provide some help but
current ideas to explain supernovae within the physics of the 1930's and 
before rely on particular initial conditions such as energy input 
via accretion from observationally unconfirmed binary partners refs.\,\cite{Feng},\cite{Perna}.    
It is clear that a new energy source such as suggested
in the susy phase transition model would be helpful.  Supersymmetry could, therefore,
be the solution to Newton's old quest for a mechanism to transform base elements
into precious metals.  A significant energy input is required to transform amy mass of 
iron into gold. 

    In the susy model it is possible to release the degeneracy energy in a small
iron core, leaving behind susy iron and other susy elements, creating 
high mass normal atoms in the outer layers, and accelerating them beyond 
their escape velocity.  As noted in ref.\,\cite{galfx} table 2,  an iron core at high
density can release an energy equivalent to $20 \%$ of its rest mass into the 
surrounding matter.  Fusion of susy nuclei could also produce elements beyond
susy iron \cite{future}, \cite{snuclear}
releasing additional energy although we neglect this possibility for the present. 

In this section we consider the endothermic reactions leading from iron 
($Z_1=26, A_1=56$) to a heavier element of atomic number, $Z_i$, and atomic weight, 
$A_i$.

The atomic mass, $m(Z,A)$ of an element with $Z$ protons and $A-Z$ neutrons is
written, in terms of constituent masses and binding energy, as
\be
     m(Z,A) = (m_p + m_e) Z + m_n (A-Z) - b(Z,A) \quad ,
\ee
with $b(Z,A)$ a relatively small nuclear and electronic binding energy.

The external energy input required to 
transform $\tilde{A}_i$ atoms of mass $m(Z_1,A_1)$ to $\tilde{A}_1$ heavier atoms of mass 
$m(Z_i,A_i)$ is
\be
    \Delta E(i) = (\tilde{A}_1 m(Z_i,A_i) - \tilde{A}_i m(Z_1,A_1)) c^2 + \delta_i
                 \quad ,
\ee
where baryon number conservation requires
\be
      \tilde{A}_i \,A_1 = \tilde{A}_1 \,A_i \quad .
\label{constraint}
\ee
The small positive correction
\be
     \delta_i \propto (\tilde{A}_i Z_1 - \tilde{A}_1 Z_i) (m_n - m_p - m_e) c^2
\ee
is the energy carried away by neutrini and photons.
The minimal $\tilde{A}_i$ and $\tilde{A}_1$ consistent with baryon number conservation are
\be
    \tilde{A}_i &=& A_i/\mathit{gcd}(A_1,A_i)\\
    \tilde{A}_1 &=& A_1/\mathit{gcd}(A_1,A_i) 
\ee
where $\mathit{gcd}(A_1,A_i)$ is the greatest common divisor of its arguments.
The degeneracy energy released in a small core of mass, $M_\mathrm{core}$,
provides the required energy
\be
     0.2 \,M_\mathrm{core}\, c^2 = \sum_i\, N_0(i)\, (\Delta E(i) +  \tilde{A}_1 \overline{K}_i )
\label{susymass}
\ee
where $N_0(i) \tilde{A}_1 $ is the number of produced atoms of element $i$.
The internal thermal energy is sufficient to bring the atoms back to their initial state before collapse while 
$\overline{K}_i$ is their additional average kinetic energy.  

The total released energy in a core collapse supernova is, observationally,
about $10^{52}\,\mathrm{ergs}$ which therefore fixes the susy core mass of 
eq.\,\ref{susymass} to be
\be
      M_\mathrm{core} \approx 2.8\,10^{-2}\,M_\odot \quad .
\label{susycore}
\ee
The total ejected mass is
\be
     M_\mathrm{Ej} = \sum_i\, N_0(i)\,\tilde{A}_1\,m(Z_i,A_i) \quad .
\ee
The relative abundance of element $i$ in the ejecta is 
proportional to $N_0(i)/\mathit{gcd}(A_1,A_i)$.
An attempt to predict the abundances in a statistical analysis is left to a future
analysis but, given that the degeneracy energy in iron is much greater than
the available fusion energy from silicon, the energy released 
in the
transition to exact susy in a small core will be sufficient to unbind the star 
apart from a 
relatively small susy remnant and create copious amounts of heavy elements.

\section{The remnant of a core collapse supernova}
\label{sec-remnant}

The remnants of core collapse supernovae have been identified in many cases
as neutron stars at progenitor masses near 10 solar masses and as
black holes of mass up to about $30\,M_\odot$ for greater progenitor masses.  
In the susy model we must address the production process of neutron stars 
and also their stability once produced.
 
In the phase transition model, an isolated neutron star like an isolated white dwarf
has a finite lifetime.  We must confirm that the predicted lifetime is not in 
disagreement with observation.  Neutron stars have a mass near $1.5\,M_\odot$ 
and a radius of about $10^{-3}\,R_E$ while white dwarfs have a comparable mass but a radius near the
Earth radius, $R_E$.

The neutron star lifetime can be estimated from eqs.\,\ref{tau} and \ref{EffVol}
using the parameters of table \,\ref{table0}.
\be
    \tau = \frac{1}{A\,  V_\mathrm{eff}} \quad .
\ee
In the case $b_0 = 0$, using the minimum lifetime, $\tau_0$ of white dwarfs from 
table\,\ref{table0}, 
\be
    \tau_{NS} \approx 10^9 \tau_0  \quad .
\label{tauNS}
\ee
This lifetime is far greater than current lifetime of the universe.

However, if $b_0 \approx 0.02$ as required to explain core collapse supernovae as
transitions to exact susy in the core, the interior of an isolated neutron star should immediately become supersymmetric. 
\be
    \tau_{NS} \approx 10^9\,\tau_0\,e^{-b_0\,\rho_\mathrm{NS}/\rho_c}\quad .
\ee
Thus, if the extension of the susy model to
core-collapse events is to be maintained and the remnant is to be identified with
putative neutron stars, these stars must be partially or totally made of
scalar neutrons.  The stability of this kind of boson star requires a strong repulsive
force at short distances. Such a repulsive hard core has been suggested even within
the standard picture in order to explain the unexpectedly high mass of neutron stars
compared to white dwarfs. The fact that quark-gluon plasmas have been suggested 
\cite{Simonetta} in the 
neutron star core testifies to the puzzles and structural uncertainties that exist 
in the theory.  If the suggestion that pulsars are, in fact, sneutron stars can be
ruled out from their observed properties, we must be prepared to abandon the susy model extension to core collapse events. 

In the susy model the prominent pulsar kicks that have been observed could be
interpreted as an off-center nucleation of a susy bubble recoiling against another
non-radiating normal or susy object.  Such pronounced kicks would be even greater
if the progenitor star had a high angular momentum.
Thus, if $b_0=0$, a normal neutron 
star would be stable and could
be produced recoiling against an unobserved susy star.  If $b_0 \approx 0.02$, as
discussed in this article,  the space within 
both the observed pulsars and recoiling non-radiating objects would be supersymmetric. 

\section{The black hole gap}
\label{sec-BHgap}

It has long been a puzzle as to why black holes are observed near the solar mass scale,
$M_\mathrm{BH} \leq 30\,M_\odot$,
and at the supermassive scale, $M_{BH} \geq 10^5\,M_\odot$ with few, if any,
at intermediate masses.
In the first work on the susy model for supernovae \cite{Biermann-Clavelli}, 
it was proposed that the gap arose because a stellar conglomeration above about
$10^5\,M_\odot$ never achieved the critical density to nucleate a susy bubble 
before becoming a black hole and, 
therefore, quietly slipped below its Schwarzschild radius.  Below the supermassive 
scale, the supernova explosion was expected to throw off $99.9\:\%$ of its mass leaving a
black hole near $30\,M_\odot$ or less.  The new considerations are:
{\flushleft 1) the time constraint:  the nucleation of a susy bubble requires not only that
critical density be achieved but also that this density persist for some time.
The transition probability is proportional to a space time volume at high density
as discussed in section \,\ref{sec-delay}.} 
{\flushleft 2) the metallicity effect: Since the degeneracy energy is proportional to
the atomic number/weight, the effective critical density is inversely proportional to 
the metallicity of the progenitor \cite{galfx}.  Thus an iron progenitor, as exists in 
core collapse supernovae will more easily produce the susy bubble than the 
lighter elements which produce type Ia supernovae.}

The net effect in the current calculation (see table\,\ref{table2} is that 
all massive stars below about $10^4\,M_\odot$ will produce prompt supernovae
dispersing large fractions of the total mass into interstellar regions.
More massive conglomerations will grow into supermassive black holes before
experiencing the susy transition.
  
An ejected mass of about 
$97\, \%$ as suggested by eq.\,\ref{susycore} will leave a black hole mass about $30\,M_\odot$ for a progenitor mass of $10^3 \,M_\odot$.
More massive stars, after falling below the Schwarzschild radius, 
will still ultimately exceed the critical density and will experience the susy transition although no matter will be ejected.  This is in accord with the result from 
string theory that supersymmetric backgrounds will explain the black hole entropy
to area ratio; see, for example, ref.\,\cite{Zwiebach}.  The suggestion from superstring theory as well as from the current work is, therefore, that all black holes exist in a supersymmetric background.
\section{Summary}
\label{sec-summary}

   In this article we have explored the possibility that the bare Coleman action 
varying as the inverse cube of the energy density is modified in such a way that
the transition probability per unit space time volume does not saturate at high 
density but continues to increase as the density grows.  Given the speculative 
aspects of the original Coleman proposal as discussed in section\,\ref{BeyondCdL}, 
the only surprising result is, perhaps, that the limit on the correction 
allowed by the fit to the delay time distribution of type Ia supernovae is 
so small ($b_0 < 0.03$).  
This allows for a unified description of both type Ia and 
core collapse supernovae.  The alternative disparate models based on the 
physics of the 1930's fail in many respects despite more than forty five years of 
theoretical and observational effort.  

Possible criticisms of the susy phase transition model that have  
arisen in discussions with colleagues are the following:

{\flushleft 1)
The Coleman formula for a phase transition through bubble nucleation was 
proposed as most likely valid in the case of large action although, even
there, not without questionable assumptions as discussed in section \ref{BeyondCdL}.
However, Coleman did not speculate as to how large the action must be for 
validity of the model nor as to how small it could be before it is 
necessarily phenomenologically untenable or uninteresting.}
{\flushleft 2)
the necessary values of the free parameters, $A$ and $\rho_c$, seem far from the
values that might be expected from fundamental particle physics.  Since every
new model has free parameters and phase transition theory often involves 
large scale collective effects this criticism does not seem necessarily damning.}

{\flushleft 3)
Neutron star physics has the potential of ruling out the susy 
model extension to core collapse supernovae although the structure of these 
stars is, at present, far from uncontroversial.  As discussed in 
section\,\ref{sec-remnant}, the existence of standard neutron stars would not be  
problematic in the bare Coleman theory with $b_0 = 0$.  However, with $b_0 = 0.02$
as required for the current extension to core collapse events,  there are strong 
implications for the structure of neutron stars as discussed in section \ref{sec-remnant}.  
If it can be shown that pulsar phenomenolgy depends on a spin 1/2 nature of its
constituents we might have to give up the susy extension.}        

The proposed unified model for supernovae adds only one additional parameter
($b_0$) to the two basic parameters of the model as originally applied to 
thermonuclear supernovae.  Besides the possible aesthetic value of a 
unified theory, the susy model addresses the standard model
challenges including the igniting of the explosion, the creation of 
sufficient numbers of high Z elements, and the black hole gap.
It would seem remarkable that, with a small departure from the bare Coleman
action, one can explain lifetimes after fusion cessation at the Gyr scale or 
longer for white dwarfs and at the scale of seconds to minutes for massive stars.
The criticisms that can be readily brought forward do not seem 
sufficiently problematic that one should dismiss from further study the 
possibility of a general susy model for supernovae. 

In the case of core collapse phenomena, the susy model can easily provide a strong
enough additional energy source to unbind the star leaving only a small 
remnant, to create the heavy elements including gold and uranium
in sufficient amounts, and to propel them
into the interstellar medium, all of which properties have defied consensus 
in standard model supernova physics for decades.  A quantitative treatment of the 
abundances of the heavy elements in the universe is left to further investigations.

The numerical results in this article are obviously approximate due, for
example, to the
replacement of volume integrals such as in eq.\,\ref{EffVol} by simpler expressions 
such as in eq.\,\ref{TauEff}.  A more precise
calculation would require knowledge of the density profile of the collapsing star
as a function of time and distance from the center.  However, it is doubtful
that such a more precise calculation would invalidate our conclusion that,
in the susy model with a non-negligible $b_0$, all 
massive stars below a certain mass will explode.

The author acknowledges useful comments from Larry Ford and Ken Olum of the Tufts Cosmology Institute. Correspondence with Peter 
Biermann of Bonn University and the University of Alabama is also acknowledged.

\end{document}